\documentclass[aps,prl,superscriptaddress,tightenlines,footinbib,longbibliography,twocolumn]{revtex4-2}
\usepackage{graphicx}
\usepackage[colorlinks,linkcolor=blue,anchorcolor=blue,citecolor=blue,urlcolor=blue]{hyperref}
\usepackage{amsmath}
\usepackage{color}
\usepackage{soul}
\usepackage{ulem}

\begin{document}

\title{Investigation of the Effect of Quantum Measurement on Parity-Time Symmetry}

\author{Wei-Chen Wang}
\affiliation{Department of Physics, College of Liberal Arts and Sciences, National University of Defense Technology, Changsha 410073, China}
\affiliation{Interdisciplinary Center for Quantum Information, National University of Defense Technology, Changsha 410073, China}

\author{Yi Xie}
\affiliation{Department of Physics, College of Liberal Arts and Sciences, National University of Defense Technology, Changsha 410073, China}
\affiliation{Interdisciplinary Center for Quantum Information, National University of Defense Technology, Changsha 410073, China}

\author{Man-Chao Zhang}
\affiliation{Department of Physics, College of Liberal Arts and Sciences, National University of Defense Technology, Changsha 410073, China}
\affiliation{Interdisciplinary Center for Quantum Information, National University of Defense Technology, Changsha 410073, China}

\author{Jie Zhang}
\affiliation{Department of Physics, College of Liberal Arts and Sciences, National University of Defense Technology, Changsha 410073, China}
\affiliation{Interdisciplinary Center for Quantum Information, National University of Defense Technology, Changsha 410073, China}

\author{Chun-Wang Wu}
\affiliation{Department of Physics, College of Liberal Arts and Sciences, National University of Defense Technology, Changsha 410073, China}
\affiliation{Interdisciplinary Center for Quantum Information, National University of Defense Technology, Changsha 410073, China}

\author{Ting Chen}
\affiliation{Department of Physics, College of Liberal Arts and Sciences, National University of Defense Technology, Changsha 410073, China}
\affiliation{Interdisciplinary Center for Quantum Information, National University of Defense Technology, Changsha 410073, China}

\author{Bao-Quan Ou}
\affiliation{Department of Physics, College of Liberal Arts and Sciences, National University of Defense Technology, Changsha 410073, China}
\affiliation{Interdisciplinary Center for Quantum Information, National University of Defense Technology, Changsha 410073, China}

\author{Wei Wu}
\affiliation{Department of Physics, College of Liberal Arts and Sciences, National University of Defense Technology, Changsha 410073, China}
\affiliation{Interdisciplinary Center for Quantum Information, National University of Defense Technology, Changsha 410073, China}

\author{Ping-Xing Chen}
\email{pxchen@nudt.edu.cn}
\affiliation{Department of Physics, College of Liberal Arts and Sciences, National University of Defense Technology, Changsha 410073, China}
\affiliation{Interdisciplinary Center for Quantum Information, National University of Defense Technology, Changsha 410073, China}

\date{\today}

\begin{abstract}
Symmetry, including the parity-time ($\mathcal{PT}$)-symmetry, is a striking topic, widely discussed and employed in many fields.  It is well-known that quantum measurement can destroy or disturb quantum systems. However, can and how does quantum measurement destroy the symmetry of the measured system? To answer the pertinent question, we establish the correlation between the quantum measurement and Floquet $\mathcal{PT}$-symmetry and investigate for the first time how the measurement frequency and measurement strength affect the $\mathcal{PT}$-symmetry of the measured system using the $^{40}\mathrm{Ca}^{+}$ ion. It is already shown that the measurement at high frequencies would break the $\mathcal{PT}$ symmetry. Notably, even for an inadequately fast measurement frequency, if the measurement strength is sufficiently strong, the $\mathcal{PT}$ symmetry breaking can occur. The current work can enhance our knowledge of quantum measurement and symmetry and may inspire further research on the effect of quantum measurement on symmetry.

\textbf{Key Word: quantum measurement, $\mathcal{PT}$-symmetry, ion trap}

\textbf{PACS number(s): 03.65.Ta, 03.65.Vf, 03.65.Yz }
\end{abstract}

\maketitle

\section{Introduction}
Symmetries determine the interactions of elementary particles \cite{002} and classify the different phases of complex systems. For nearly a decade, a striking discovery revealed that the parity-time ($\mathcal{PT}$)-symmetric Hamiltonians, despite their non-Hermitian nature, can have real eigenvalues \cite{PTReview0,PTReview1,PTReview2,PTReview3,PTReview4,PTReview5}. By controlling the parameters of the $\mathcal{PT}$-symmetric system, spontaneous $\mathcal{PT}$ symmetry breaking can occur at an exceptional point (EP) \cite{EPReview01,EPReview02}, where both the eigenvalues and eigenstates of the system coalesce. Thus, there are two phases, one being the spontaneous $\mathcal{PT}$ symmetry breaking phase (PTBP), another being the spontaneous $\mathcal{PT}$ symmetry unbreaking phase (PTSP). Recently, $\mathcal{PT}$-symmetric systems have been successfully implemented in classical optical systems \cite{SML1,LIL,EPSensing1} and quantum systems by exploiting the properties of open quantum systems \cite{PTQ,024,Tomo,2021wang,2013zhang}. This has laid the experimental foundation for further research on quantum $\mathcal{PT}$-symmetry. Compared with static time-independent $\mathcal{PT}$-symmetric systems, time-dependent Floquet $\mathcal{PT}$-symmetric systems \cite{PTRabi,PTRabi02,2019Li} have attracted greater attention.

Quantum measurement is a lasting research topic whose history dates back to the birth of quantum mechanics. However, the physical mechanisms of quantum measurement still contain some unknowns. The mathematical formalism of the quantum measurement was introduced by von Neumann \cite{001}. From a physical point of view, measurement is the interaction between the measured system and an external system, thus playing the role of a measuring apparatus. Measurement frequency and strength are two important features of the measurement. The evolution of the measured system is hindered (even stopped) when the measurement frequency increases from small to large, a well-known quantum Zeno effect \cite{Peres1998Zeno,1977TheZeno,1977Time,NAKAZATO1996}. Moreover, based on the strength of quantum measurement, two complementary measurement schemes have been widely investigated in various quantum systems: von Neumann's projective 'strong' measurement and Aharonov's weak measurement \cite{1988Aharonov,2019wu}. The quantum weak measurement can be used to achieve an ultrasensitive measurement \cite{weakm01,weakm02}. The recent discovery of weak–to–strong transition of quantum measurement in a trapped-ion system \cite{2020Weak} has stimulated further research into the mechanisms of quantum measurements.

How quantum measurement affects the symmetry of a quantum system is a fundamental question. Some previous studies Ref.\cite{QZ/AZ02} are related to this topic, but their conclusion remains far from the actual answer. Here, we design a quantum Floquet $\mathcal{PT}$-symmetric system adapted to the existing quantum measurement model by using the trapped-ion system, where in the frequency and strength of the measurement can be modulated conveniently. We analyze the effect of measurement frequency and strength on the $\mathcal{PT}$ symmetry and observe that the measurement at high frequencies would break the $\mathcal{PT}$ symmetry; as long as the measurement strength is strong enough, the $\mathcal{PT}$ symmetry breaking can occur even if the measurement frequency is not fast enough. This work may improve the existing knowledge of quantum measurement and symmetry and inspire further research in this area.

\begin{figure*}[htb]
	\centering
	{
		\includegraphics[width= 0.92\textwidth]{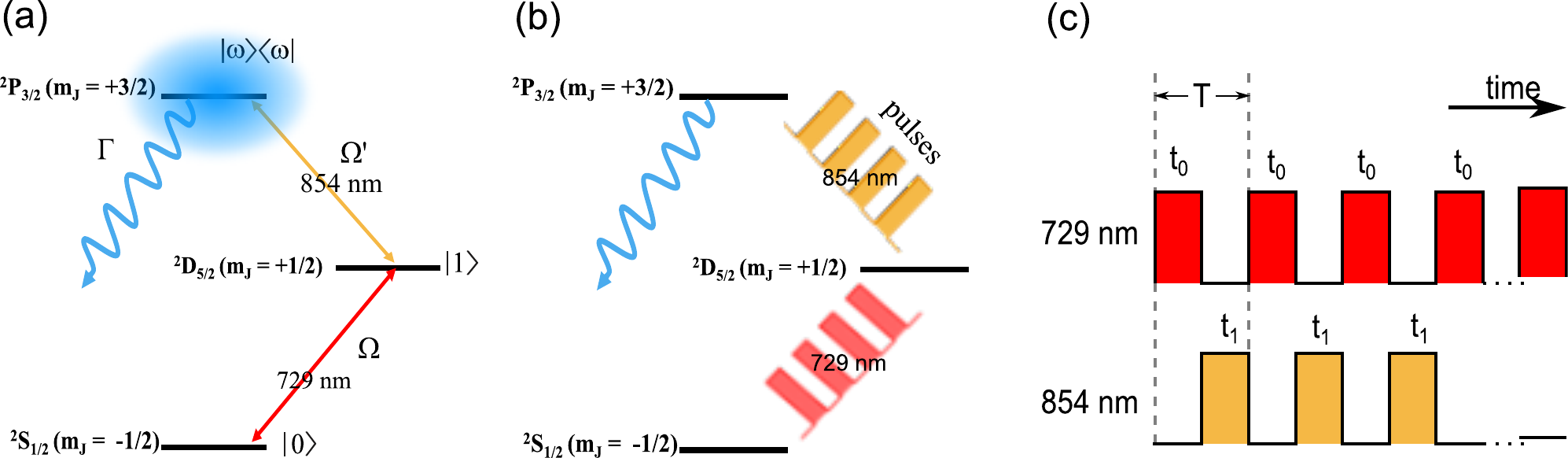}
	}
    \caption{\label{Fig1}(Color online) The experimental and theoretical models of quantum measurement (a) The energy levels of the $^{40}\mathrm{Ca}^{+}$ ion with the internal states $|0\rangle$, $|1\rangle$, and $|\omega\rangle$ corresponding to the energy levels $^{2}\mathrm{S}_{1/2}(m_{J}=-1/2)$, $^{2}\mathrm{D}_{5/2}(m_{J}=+1/2)$, and $^{2}\mathrm{P}_{3/2}(m_{J}=+3/2)$ with a flat continuum photon field, respectively. (b) An experimental scheme for the pulse measurement is proposed, based on the internal states of $^{40}\mathrm{Ca}^{+}$. The 854\,nm laser controls the measurement process, while the 729\,nm laser controls the evolution between the internal states of $^{40}\mathrm{Ca}^{+}$. (c) The pulse time sequence of laser control. The coupling strength $\Omega$ between the internal states is controlled by the intensity of the 729\,nm laser, and the evolution time $t_{0}$ is controlled by the duration time of the 729\,nm laser. The interaction strength $\gamma$ is controlled by the intensity of the 854\,nm laser, and the measurement time $t_{1}$ is controlled by the duration of the 854\,nm laser.}
\end{figure*}

\section{Theoretical and experimental schemes}
Let us now consider a simple scheme, as shown in FIG.~\ref{Fig1}(a). In the $^{40}\mathrm{Ca}^{+}$ system, the level $^{2}\mathrm{S}_{1/2}(m_{J}=-1/2)$ is set to $|0\rangle$, and the metastable level $^{2}\mathrm{D}_{5/2}(m_{J}=+1/2)$ is set to $|1\rangle$. $^{2}\mathrm{S}_{1/2}(m_{J}=-1/2)$ and $^{2}\mathrm{D}_{5/2}(m_{J}=+1/2)$ can be coupled by the 729\,nm laser, and the coupling strength is determined by the laser intensity. Electrons can be excited from $^{2}\mathrm{D}_{5/2}(m_{J}=+1/2)$ to level $^{2}\mathrm{P}_{3/2}(m_{J}=+3/2)$ by the 854\,nm laser. Since the natural linewidth of $^{2}\mathrm{P}_{3/2}(m_{J}=+3/2)$ is large, i.e., the lifetime of $^{2}\mathrm{P}_{3/2}(m_{J}=+3/2)$ is very short, photons drain the population away from $^{2}\mathrm{P}_{3/2}(m_{J}=+3/2)$ rapidly (toward another level not shown in the figure). When the 854\,nm laser is turned on, if the electron is in $^{2}\mathrm{D}_{5/2}(m_{J}=+1/2)$, it gets excited to $^{2}\mathrm{P}_{3/2}(m_{J}=+3/2)$ and spontaneously emits photons. Theoretically, the process can be regarded as a measurement of $^{2}\mathrm{D}_{5/2}(m_{J}=+1/2)$, whether or not the photon emitted spontaneously from $^{2}\mathrm{P}_{3/2}(m_{J}=+3/2)$, is detected. Therefore, according to the scheme given in \cite{Facchi2009}, the Hamiltonian is composed of the internal states $|0\rangle$ and $|1\rangle$, and the photon field can be written as:
\begin{eqnarray}
	H_{total} &=& \frac{\Omega}{2}(|0\rangle\langle 1|+|1\rangle\langle 0|) \nonumber \\
	&+& \int \mathrm{d}\omega \omega |\omega\rangle\langle \omega| + \sqrt{\frac{\gamma}{\pi}}\int \mathrm{d}\omega(|\omega\rangle\langle 1|+|1\rangle\langle \omega|).\label{eq1}
\end{eqnarray}
The photon field and $^{2}\mathrm{P}_{3/2}(m_{J}=+3/2)$ constitute the measuring apparatus, described by ${|\omega\rangle\langle \omega|}$, and the measured system is composed of the internal states $|0\rangle$ and $|1\rangle$, wherein $\Omega$ is the coupling strength between $|0\rangle$ and $|1\rangle$. Moreover, $\gamma$ is the interaction strength between the measuring apparatus and the measured system \cite{Facchi2009}, determined by the intensity of the 854\,nm laser and the natural linewidth of $^{2}\mathrm{P}_{3/2}(m_{J}=+3/2)$. Solely focusing on the dynamic evolution of the measured system, the Hamiltonian $H_{total}$ can be reduced to an effective Hamiltonian $H_{eff}$,
\begin{eqnarray}
	H_{eff} = -i \gamma |1\rangle\langle 1| + \frac{\Omega}{2} ( |0\rangle\langle 1| + |1\rangle\langle 0|).  \label{eq2}
\end{eqnarray}
This relation yields the Rabi oscillations of frequency $\Omega/2$, but at the same time absorbs the population of $|1\rangle$, thereby performing a "measurement." Thus, using the $^{40}\mathrm{Ca}^{+}$ system, we design a continuous measurement model \cite{1988Milburn,1998Continuous,BALZER2000,2006Continuous} wherein the 729\,nm laser enables the evolution of the quantum states, while the 854\,nm laser performs the measurements.

Now, let us rewrite Eq.~(\ref{eq2}) as $H_{eff}=H_{\mathcal{PT}}-\frac{i\gamma}{2} \mathbf{I}$, where $H_{\mathcal{PT}} = \frac{\Omega}{2} \sigma_x - i\frac{\gamma}{2} \sigma_z$ is the $\mathcal{PT}$-symmetric Hamiltonian with balanced gain and loss, $\sigma_{x(z)}$ is the Pauli matrix, and $\mathbf{I}$ is the identity operator. This continuous measurement model corresponds to the static passive $\mathcal{PT}$-symmetric system \cite{2021wang}, which is time-independent. By solving the eigenvalue of the Hamiltonian $H_{\mathcal{PT}}$, the expression of the discriminant of the $\mathcal{PT}$ symmetry is obtained as follows:
\begin{eqnarray}\label{eq2-1}
	\left\{
	\begin{aligned}
		& \frac{\gamma}{\Omega} < 1, && \textit{PTSP}\ ,  \\
		& \frac{\gamma}{\Omega} > 1, && \textit{PTBP}\ .
	\end{aligned}
	\right.
\end{eqnarray}

However, the continuous measurement model cannot define the measurement frequency and strength well. Hence, we must consider a pulse measurement model \cite{1989peres,1990peres,Cook1988,1990ZenoIon}. The experimental scheme suitable for $^{40}\mathrm{Ca}^{+}$ system is shown in FIG.~\ref{Fig1}(b). In this scheme, instead of constantly interacting with ions, the laser is divided into pulses, and the pulse time sequence is shown in FIG.~\ref{Fig1}(c). The system is first evolved under the 729\,nm laser drive for $t_{0}$. Then, the 729\,nm laser is switched off, and the 854\,nm laser is switched on to measure the quantum state, with the duration time of $t_{1}$, and so on. In addition to controlling the pulse duration, one can also control the pulse intensity. Under the action of 729\,nm and 854\,nm laser pulses, the quantum state alternately performs the "evolution-measurement," which is the pulse measurement model of $^{40}\mathrm{Ca}^{+}$ system.

 Referring to the Hamiltonian $H_{eff}$ of the continuous measurement model, the Hamiltonian $H_{eff}(t)$ of the pulse measurement model can be written as:
\begin{eqnarray}\label{eq3}
 &H_{eff}&(t) = -i \gamma(t) |1\rangle\langle 1| + \frac{\Omega(t)}{2} ( |0\rangle\langle 1| + |1\rangle\langle 0|) \nonumber \\
 &H_{eff}&(t) = H_{\mathcal{PT}}(t)-\frac{i\gamma(t)}{2}\mathbf{I} \\
    & &\left\{
	\begin{aligned}
		& \Omega(t) = \Omega, \gamma(t)=0, &&0< t \leq t_{0},  \nonumber\\
		&  \Omega(t)= 0, \gamma(t) = \gamma, && t_{0}< t \leq t_{0}+t_{1},\nonumber\\
	\end{aligned}
	\right. 
\end{eqnarray}
  Where $H_{\mathcal{PT}}(t) = \frac{\Omega(t)}{2} \sigma_x - i\frac{\gamma(t)}{2} \sigma_z$. $H_{eff}(t)$ is a periodic time-dependent Hamiltonian, and $H_{eff}(t+T)=H_{eff}(t)$, where $T=t_{0}+t_{1}$ is the period of the Hamiltonian $H_{eff}(t)$. Notably,
 $H_{\mathcal{PT}}(t)$ is no longer a static $\mathcal{PT}$-symmetric system. According to Eq.~(\ref{eq2-1}), the $\mathcal{PT}$ symmetry of $H_{\mathcal{PT}}(t)$ alternates with time.
 The $\mathcal{PT}$ symmetry of $H_{\mathcal{PT}}(t)$ becomes difficult to describe, and the evolution characteristics become ambiguous. Therefore, we will describe the $\mathcal{PT}$ symmetries of such time-dependent systems using the Floquet theorem \cite{PTRabi,PTRabi02}.

\section{Results and discussion}
According to the Hamiltonian in Eq.~(\ref{eq3}), the operator $U(T,0)$ of an "evolution-measurement" can be written as:
\begin{eqnarray}
	U(T,0) = \begin{bmatrix} e^{-\gamma t_{1}} cos(\Omega t_{0}/2) & -i e^{-\gamma t_{1}} sin(\Omega t_{0}/2) \\ -i sin(\Omega t_{0}/2) & cos(\Omega t_{0}/2) \end{bmatrix}.  \label{eq4}
\end{eqnarray}
 Here, $\Omega t_{0}/2$ is the measurement interval related to the measurement frequency. We set the initial state $\rho(0)$ of the measured system as $|0\rangle\langle 0| $. After applying the $U(T,0)$ $n$ times, the probability that the measured system is still in the initial state becomes
\begin{eqnarray}
	P_{0} &=& |\langle 0|U^{n}(T,0)\rho (0)U^{\dagger n}(T,0)|0\rangle|^{2}.    \label{eq4-1}
\end{eqnarray}
where $P_{0}$ is the survival probability. The expression of $U(T,0)$ is similar to Eq.(47) in the literature \cite{1989peres}. When $e^{-\gamma t_{1}} \rightarrow 1$, the evolution of the measured system is consistent with the general unperturbed Rabi oscillation, the measurement is fuzzy, and the wave packet of the measuring apparatus is not separated \cite{1989peres}. When $e^{-\gamma t_{1}} \rightarrow 0$, literature \cite{1989peres} describes that the measurement is very strong, and the wave packet of the measuring apparatus becomes sharp. As discussed above, $\gamma t_{1}$ is related to the separation distance of the wave packet of the measuring apparatus. Therefore, $\gamma t_{1}$ can be defined as the measurement strength, and an increase in $\gamma t_{1}$ implies that the measurement strength changes from weak to strong. Finally, from a physical point of view, the operator $U(T,0)$ can be described as follows: the measured system evolves $t_0 $with the driving strength $\Omega/2$ and is subsequently measured with the strength $\gamma t_{1}$.

To discuss the $\mathcal{PT}$ symmetry of the pulse measurement model, we consider the Floquet Hamiltonian $\mathcal{H}_{F}(t)=H_{eff}(t)-i \partial_{t}$ \cite{PTRabi,PTRabi02}. By calculating the eigenvalue of $\mathcal{H}_{\mathcal{PT}}=H_{\mathcal{PT}}(t)-i \partial_{t}$ \cite{2003Floquet}, we can obtain the expression for the discriminant of $\mathcal{PT}$ symmetry of the measured system (detailed derivation is in the appendix),
\begin{eqnarray}\label{eq5}
	\left\{
	\begin{aligned}
		& cos^{2}(\frac{\Omega t_{0}}{2}) cosh^{2}(\frac{\gamma t_{1}}{2}) < 1, && \textit{PTSP},  \\
		&  cos^{2}(\frac{\Omega t_{0}}{2}) cosh^{2}(\frac{\gamma t_{1}}{2}) > 1, && \textit{PTBP}.
	\end{aligned}
	\right.
\end{eqnarray}
When $cos^{2}(\frac{\Omega t_{0}}{2}) cosh^{2}(\frac{\gamma t_{1}}{2})=1$, we obtain the EP of the measured system. Eq.~(\ref{eq5}) indicates that the $\mathcal{PT}$ symmetry of the measured system is determined by both the measurement interval $\Omega t_{0}$ and strength $\gamma t_{1}$.

\begin{figure*}[htbp]
	\centering
	{
		\includegraphics[width= 0.93\textwidth]{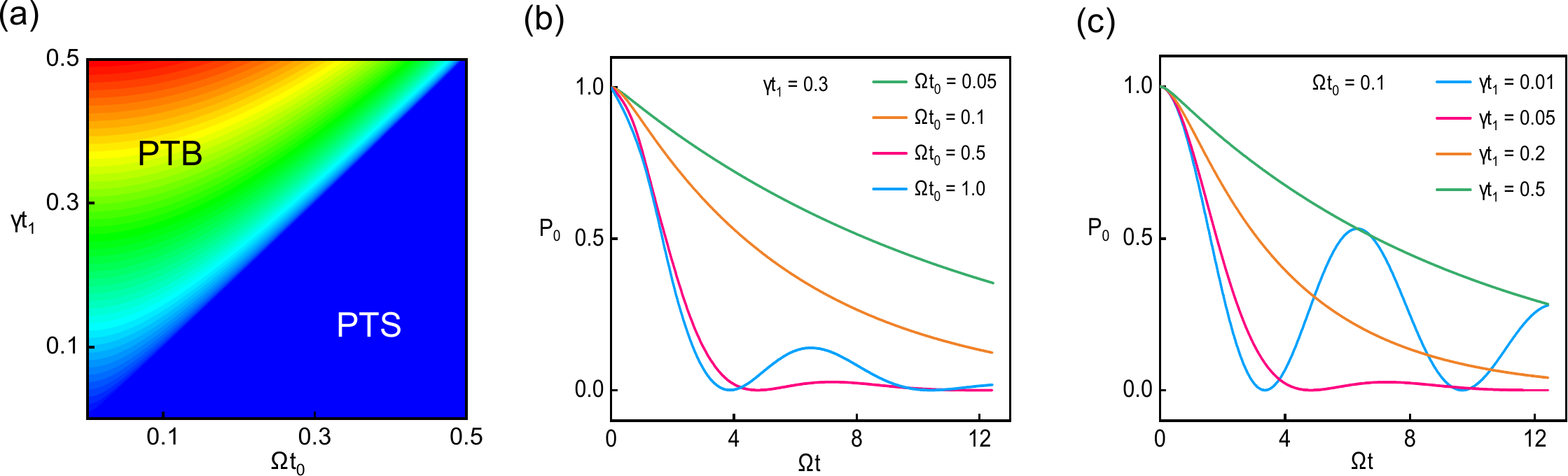}
	}
	\caption{\label{Fig2}(Color online) (a) PT phase diagram of the measured system in the Floquet representation. (b) The survival probability $P_{0}$ when the measurement strength is fixed, $\gamma t_{1}=0.3$, and the measurement interval $\Omega t_{0}=0.05,0.1,0.5,1.0$. (c) The survival probability $P_{0}$ when the measurement interval is fixed, $\Omega t_{0}=0.1$, and the measurement strength $\gamma t_{1}=0.01,0.05,0.2,0.5$.}
\end{figure*}
Thus, we establish a correlation between the general quantum measurement model and the Floquet $\mathcal{PT}$-symmetry system. Next, we investigate the effects of measurement interval and measurement strength on the $\mathcal{PT}$ symmetry. According to Eq.~(\ref{eq5}), FIG.~\ref{Fig2}(a) shows the $\mathcal{PT}$ symmetric phase diagram of the measured system with respect to the measurement interval $\Omega t_{0}$ and measurement strength $\gamma t_{1}$. By choosing different $\Omega t_{0}$ and $\gamma t_{1}$, we can obtain different symmetry phases of the measured system.

We set the initial state of the measured system as $|0\rangle$. The survival probability of the measured system is shown in FIG.~\ref{Fig2}(b), when fixing the measurement strength ($\gamma t_{1}=0.3$). The measured system is in the PTSP, when $\Omega t_{0}=0.5,1$. When $\Omega t_{0}=0.05,0.1$, the measured system is in PTBP. This implies that when the measurement frequency is fast enough, the coherent evolution between the internal states disappears, and the measured system changes from PTSP to PTBP. When $\Omega t_{0}$ is further reduced, the evolution of the measured system is significantly slower in the PTBP. Therefore, the measurement at a high frequency breaks the $\mathcal{PT}$ symmetry of the measured system. The dynamic characteristic of the unbroken $\mathcal{PT}$ symmetry system is the oscillation and that of the broken $\mathcal{PT}$ symmetry system is an exponential decrease in the survival probability.

As shown in FIG.~\ref{Fig2}(c), when the measurement frequency is fixed ($\Omega t_{0}=0.1$), and the measurement strength $\gamma t_{1}$ is changed from weak to strong, the evolution of the measured system will also change significantly. When $\gamma t_{1}=0.01,0.05$, the measured system is in the PTSP. The evolution of the measured system is oscillation, and there are still coherent transitions between the internal states. When $\gamma t_{1}=0.2,0.5$, the strong measurement strength breaks the $\mathcal{PT}$-symmetry of the measured system. The evolution of the measured system becomes slower as the measurement strength becomes stronger in the PTBP. Therefore, we find that even if the measurement frequency is not fast enough, as long as the measurement strength is strong enough, the $\mathcal{PT}$ symmetry of the measured system will be broken.
\begin{figure}[htb]
	\centering
	{
		\includegraphics[width= 0.44\textwidth]{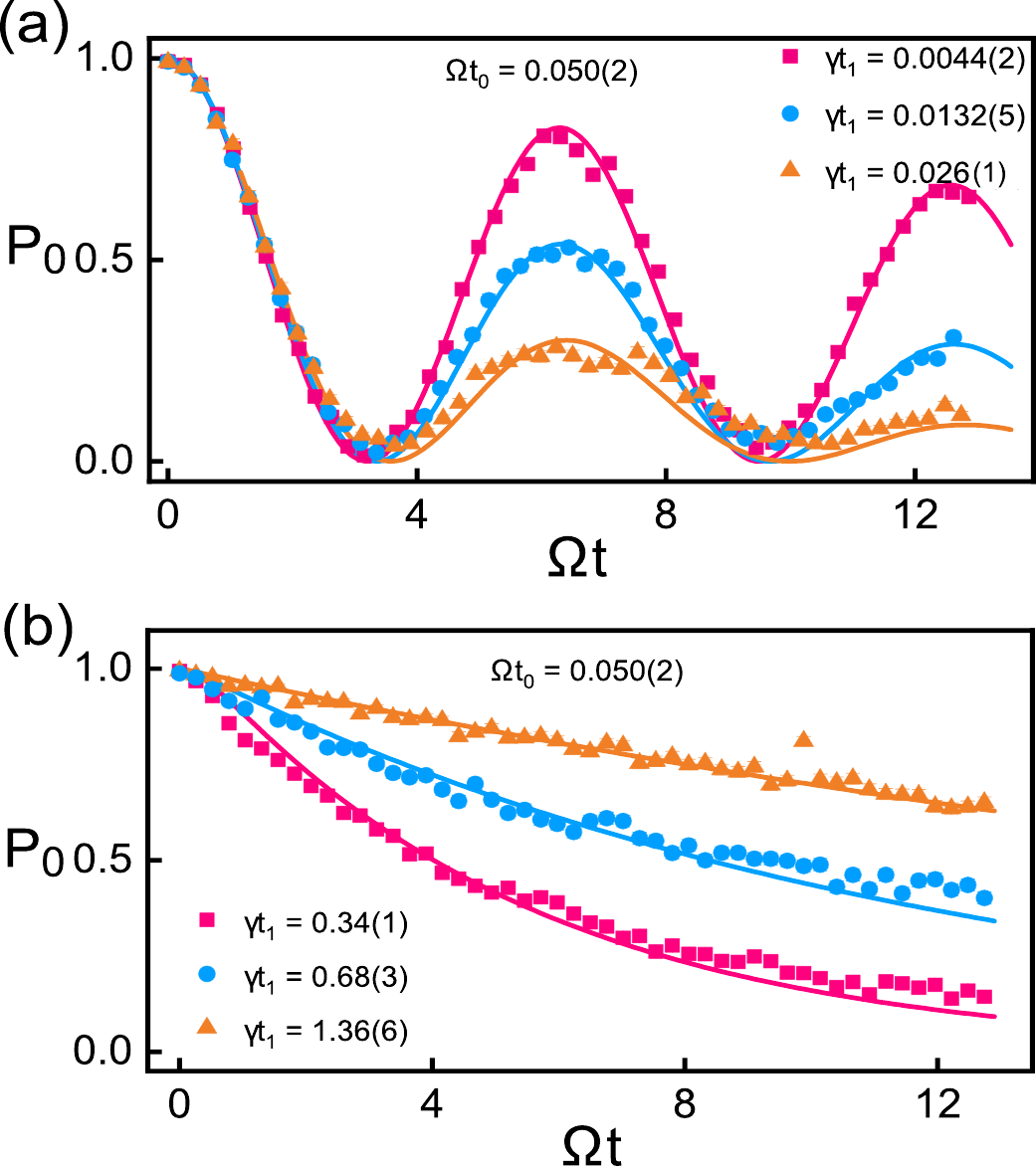}
	}
	\caption{\label{Fig3}(Color online) (a) The survival probability $P_{0}$ when the measurement interval is fixed, $\Omega t_{0}=0.050(2)$, and the measured system is in PTSP. (b) The survival probability $P_{0}$ when the measurement interval is fixed, $\Omega t_{0}=0.050(2)$, and the measured system is in the PTBP. The error bars represent the standard deviation of the measurements.}
\end{figure}

Next, we experimentally observe the effect of the measured strength on the $\mathcal{PT}$ symmetry. We set the measurement interval as $\Omega t_{0}=0.050(2)$ and the initial state as $|0\rangle$. The dynamic evolution of the measured system is shown in FIG.~\ref{Fig3}(a). For the measurement strength $\gamma t_{1}=0.0044(2),0.0132(5),0.026(1)$, the measured system is in PTSP. In FIG.~\ref{Fig3}(b), $\gamma t_{1}=0.34(1),0.68(3),1.36(6)$; the measured system is in PTBP. The dynamic evolution of the measured system is no longer an oscillation but an exponential decay. When the measured system is in PTBP, with the increase in the measurement strength $\gamma t_{1}$, the transition rate from $|0\rangle$ to $|1\rangle$ becomes slower.

\begin{figure}[htb]
	\centering
	{
		\includegraphics[width= 0.43\textwidth]{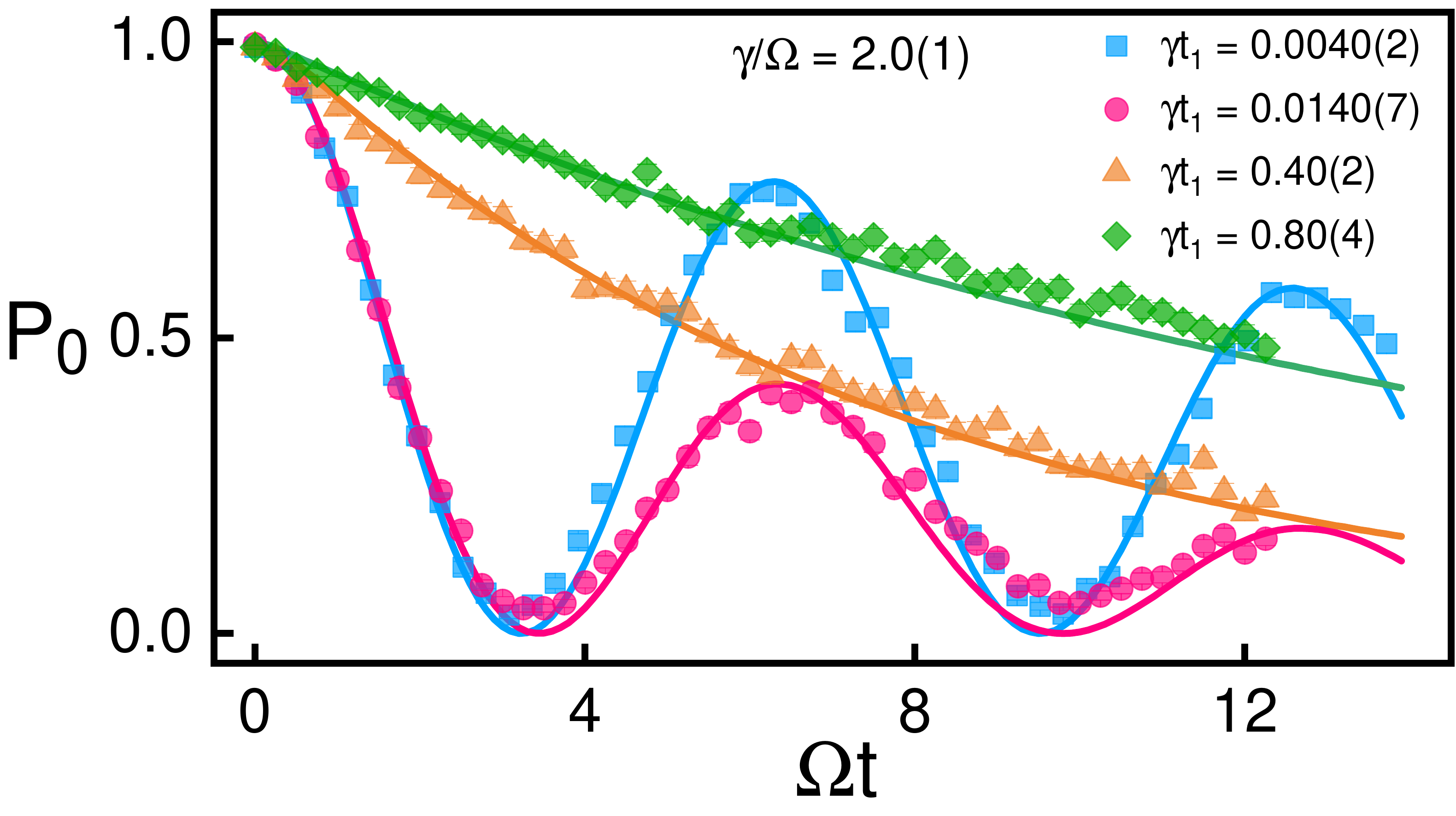}
	}
	\caption{\label{Fig4}(Color online) The survival probability $P_{0}$ when $\Omega t_{0}=0.050(2)$ and $\gamma/\Omega=2.0(1)$. When $\gamma/\Omega=2$, the pulse measurement model can still be in PTSP.}
\end{figure}

Interestingly, although both static $\mathcal{PT}$-symmetry system and Floquet $\mathcal{PT}$-symmetry system satisfy the $\mathcal{PT}$-symmetry, the expression of the discriminant Eq.~(\ref{eq5}), obtained by solving the eigenvalues of $\mathcal{H}_{\mathcal{PT}}(t)$, is quite different from Eq.~(\ref{eq2-1}). Since $\mathcal{H}_{\mathcal{PT}}(t)$ is time-dependent, Eq.~(\ref{eq5}) depends not only on $\gamma$ and $\Omega$, but also on time $t_0$ and $t_1$.
Hence, we experimentally investigated the difference between the Floquet $\mathcal{PT}$-symmetric system and static passive $\mathcal{PT}$-symmetric system. According to Eq.~(\ref{eq2-1}), when $\gamma/\Omega=2$, the static $\mathcal{PT}$-symmetric system is in PTBP and the dynamic evolution of the system should no longer oscillate. For the Floquet $\mathcal{PT}$-symmetric system, the experimental results are demonstrated in FIG.~\ref{Fig4}. We set the initial state as $|0\rangle$. When $\Omega t_{0}=0.050(2)$, $\gamma t_{1}=0.0040(2),0.0140(7)$, the evolution of the measured system is still under oscillation. Therefore, the $\mathcal{PT}$ symmetry is not broken. When $\gamma t_{1}=0.40(2),0.80(4)$, the evolution of the measured system is exponential, and the $\mathcal{PT}$ symmetry is broken. The experimental results do not agree with the conclusion given by Eq.~(\ref{eq2-1}) but agree with the conclusion given by Eq.~(\ref{eq5}).

\section{Conclusion}
We probe the effect of quantum measurement on the symmetry of the measured system and establish the correlation between the quantum measurement and the Floquet $\mathcal{PT}$-symmetry. Additionally, an experimental scheme to investigate the $\mathcal{PT}$ symmetry of the measured system was designed by the trapped-ion system. Using Floquet theorem, we obtain the expression of the discriminant of $\mathcal{PT}$ symmetry of the measured system, which is very different from that of the static $\mathcal{PT}$-symmetric system. We observe that the measurement interval $\Omega t_{0}$ and strength $\gamma t_{1}$ affect the $\mathcal{PT}$ symmetry of the measured system. When the measurement strength $\gamma t_{1}$ is fixed and the measurement frequency is fast enough, the $\mathcal{PT}$ symmetry of the measured system is broken, and the evolution of the system is hindered by the measurement. Conversely, when the measurement interval $\Omega t_{0}$ is fixed, the sufficiently strong measurement will break the $\mathcal{PT}$ symmetry of the measured system.

There are some open questions to be investigated further. What is the physical mechanism of $\mathcal{PT}$ symmetry breaking caused by quantum measurement? Can the measurement affect the other symmetry?

\section{Acknowledgments}
This work is supported by the National Basic Research Program of China under Grant No.2016YFA0301903, the National Natural Science Foundation of China under Grant.12074433, 12004430, 12174447, 12174448, 11904402.

\nocite{*}

\bibliography{ZenoPTmanuscriptV7}

\appendix
\setcounter{equation}{0}
\renewcommand{\theequation}{A\arabic{equation}}
\setcounter{figure}{0}
\renewcommand{\thefigure}{A\arabic{figure}}

\section{Appendix}
\subsection{A1. Discrimination of PT symmetry}
First, according to the properties of $\mathcal{PT}$-symmetry, several methods can be used to determine whether the system is in the $\mathcal{PT}$ symmetry phase (PTSP) or $\mathcal{PT}$ symmetry breaking phase (PTBP). For example, the Hamiltonian $H_{\mathcal{PT}}$ and $\mathcal{PT}$ operator have common eigenstates in PTSP, while no common eigenstates in PTBP. The second method is a widely accepted method, which involves the calculation of the eigenvalue of the Hamiltonian $H_{\mathcal{PT}}$.

Next, we discuss how to determine the symmetry of Floquet $\mathcal{PT}$-symmetry system. First, we introduce the Floquet method. Floquet method can be used to solve the periodic time-dependent Hamiltonian; for example, the Schrodinger equation of the time-dependent Hamiltonian $H(t)$ is:
\begin{eqnarray}\label{aeq1}
	H\left(t\right)\Psi\left(t\right)=i\hbar\frac{\partial}{\partial t}\Psi\left(t\right).
\end{eqnarray}
If $H(t)=H(t+T)$, where T is the period. Then, according to the Floquet's theory, there are solutions to Eq.~(\ref{aeq1})
\begin{eqnarray}\label{aeq3}
	\Psi_\alpha\left(t\right)=\exp{\left(-\frac{i\epsilon_\alpha t}{\hbar}\right)}\Phi_\alpha\left(t\right), 
\end{eqnarray}
where $\Psi_\alpha\left(t\right)$ is called the Floquet state; $\Phi_\alpha\left(t\right)=\Phi_\alpha(t+T)$ satisfies periodicity and is known as the Floquet mode; $\epsilon_\alpha$ is a constant number that does not vary with time and is defined as a quasi-energy level. Hence, given $\Phi_\alpha\left(t\right)$ and $\epsilon_\alpha$ for a particular $H\left(t\right)$, the solution of the wave function $\Psi\left(t\right)$ at any time $t$ can be obtained as:
\begin{eqnarray}\label{aeq4}
	\Psi\left(t\right)&=&\sum_{\alpha} c_\alpha\Psi_\alpha\left(t\right) \nonumber\\
	&=&\sum_{\alpha} c_\alpha\exp{\left(-\frac{i\epsilon_\alpha t}{\hbar}\right)}\Phi_\alpha\left(t\right),
\end{eqnarray}
where the coefficient $c_\alpha$ can be obtained from the initial wave function $\Psi\left(0\right)=\sum_{\alpha} c_\alpha\Psi_\alpha\left(0\right)$. 
By substituting Eq.~(\ref{aeq3}) into Eq.~(\ref{aeq1}), the eigen-equations about Floquet mode and quasi-energy can be obtained as:
\begin{eqnarray}\label{aeq5}
	\mathcal{H}\left(t\right)\Phi_\alpha\left(t\right)=\epsilon_\alpha\Phi_\alpha\left(t\right),
\end{eqnarray}
where $\mathcal{H}\left(t\right)=H\left(t\right)-i\hbar\frac{\partial}{\partial t}$ is the Floquet Hamiltonian.

The evolution of the wave function $\Psi\left(t\right)$ satisfies the following form:
\begin{eqnarray}\label{aeq6}
	U\left(T+t,t\right)\Psi\left(t\right)=\Psi\left(T+t\right).
\end{eqnarray}
Substitute Eq.~(\ref{aeq4}) into Eq.~(\ref{aeq6}) to obtain
\begin{eqnarray}\label{aeq7}
	&U\left(T+t,t\right)& \exp{\left(-\frac{i\epsilon_\alpha t}{\hbar}\right)}\Phi_\alpha\left(t\right) \nonumber\\
	&=&\exp{\left(-\frac{i\epsilon_\alpha(T+t)}{\hbar}\right)}\Phi_\alpha\left(T+t\right).
\end{eqnarray}
Since $\Phi_\alpha\left(t\right)=\Phi_\alpha(t+T)$, the formula above can be obtained
\begin{eqnarray}\label{aeq8}
	U\left(T+t,t\right)\Phi_\alpha\left(t\right)&=&\exp{\left(-\frac{i\epsilon_\alpha T}{\hbar}\right)}\Phi_\alpha\left(t\right) \nonumber\\
	&=&\eta_\alpha\Phi_\alpha\left(t\right), 
\end{eqnarray}
which shows that the Floquet modes are the eigenstates of the one-period propagator.
Therefore, we can find the Floquet modes and quasi-energies $\epsilon_\alpha=-\hbar \ln (\eta_\alpha)/T$ by numerically calculating $U\left(T+t,t\right)$ and diagonalizing it.

As discussed above, Floquet $\mathcal{PT}$-symmetric Hamiltonian can be written as $\mathcal{H}_{\mathcal{PT}}\left(t\right)=H_{\mathcal{PT}}\left(t\right)-i\hbar\frac{\partial}{\partial t}$, where $H_{\mathcal{PT}}\left(t\right)$ is the $\mathcal{PT}$-symmetric Hamiltonian that varies periodically with time. To evaluate the symmetry of the Floquet $\mathcal{PT}$-symmetric system, the eigenvalue $\epsilon_i$ of $\mathcal{H}_{\mathcal{PT}}\left(t\right)$ can be calculated. Since $\epsilon_i$ is related to the eigenvalue $\eta_i$ of the time evolution operator $U\left(T+t,t\right)$, the expression of discriminant of the symmetry of Floquet $\mathcal{PT}$-symmetric system can be obtained. When $\eta_i$ is imaginary, $\epsilon_i$ is real and the system is in PTSP; when $\eta_i$ is real, $\epsilon_i$ is imaginary and the system is in PTBP.

In this paper, the Hamiltonian of Floquet $\mathcal{PT}$-symmetric system is as follows:
\begin{eqnarray}\label{aeq9}
	H_{eff}&(t)& = -i \gamma(t) |1\rangle\langle 1| + \frac{\Omega(t)}{2} ( |0\rangle\langle 1| + |1\rangle\langle 0|) \\
	& &\left\{
	\begin{aligned}
		& \Omega(t)\neq 0, \gamma(t)=0, &&0< t \leq t_{0},  \nonumber\\
		&  \Omega(t)= 0, \gamma(t)\neq0, && t_{0}< t \leq t_{0}+t_{1},\nonumber\\
	\end{aligned}
	\right. 
\end{eqnarray}
The time evolution operator within a period $T=t_0+t_1$ is given by: 
\begin{eqnarray}\label{aeq10}
	& &U(T,0)=e^{-\gamma|1\rangle\langle 1| t_1}e^{-i\frac{\mathrm{\Omega}}{2}(|1\rangle\langle 0|+|0\rangle\langle 1|)t_0}  \nonumber\\ 
	&=& \left[\begin{matrix}exp(-\gamma t_{1})cos(\frac{\mathrm{\Omega}t_{0}}{2})&-iexp(-\gamma t_{1})sin(\frac{\mathrm{\Omega}t_{0}}{2})\\-isin(\frac{\mathrm{\Omega}t_{0}}{2})&cos(\frac{\mathrm{\Omega}t_{0}}{2})\\\end{matrix}\right].
\end{eqnarray}
The eigenvalues of $U(T,0)$ are solved as follows:
\begin{eqnarray}\label{aeq11}
	\eta_\pm &=& \frac{1}{2}e^{-\gamma t_1}[cos(\frac{\mathrm{\Omega}t_0}{2})+e^{\gamma t_1}cos(\frac{\Omega t_0}{2})]  \nonumber\\
	&\pm& 2e^{-\frac{1}{2}\gamma t_1} [\sqrt{cosh^{2}(\frac{\gamma t_{1}}{2})cos^{2}(\frac{\Omega t_0}{2})-1}],
\end{eqnarray}
From the above formula, $\sqrt{cosh^{2}(\frac{\gamma t_{1}}{2})cos^{2}(\frac{\Omega t_0}{2})-1}$ determines whether $\eta_\pm$ is complex.
Then, the discriminant of symmetry in the text is obtained as follows:
\begin{eqnarray}\label{aeq12}
	\left\{
	\begin{aligned}
		& cos^{2}(\frac{\Omega t_{0}}{2}) cosh^{2}(\frac{\gamma t_{1}}{2}) < 1, && \textit{PTSP}\,  \\
		& cos^{2}(\frac{\Omega t_{0}}{2}) cosh^{2}(\frac{\gamma t_{1}}{2}) > 1, && \textit{PTBP}\ .
	\end{aligned}
	\right.
\end{eqnarray}
In fact, it is more rigorous to solve the eigenvalue of $\mathcal{H}_{PT}\left(t\right)=H_{PT}\left(t\right)-i \partial_{t}$. However, the expression of the discriminant of $\mathcal{PT}$ symmetry obtained by solving $\mathcal{H}_{F}(t)$ or $\mathcal{H}_{PT}$ is the same.

\subsection{A2. Floquet PT symmetric systems}
\begin{figure}[htb]
	\centering
	{
		\includegraphics[width= 0.35\textwidth]{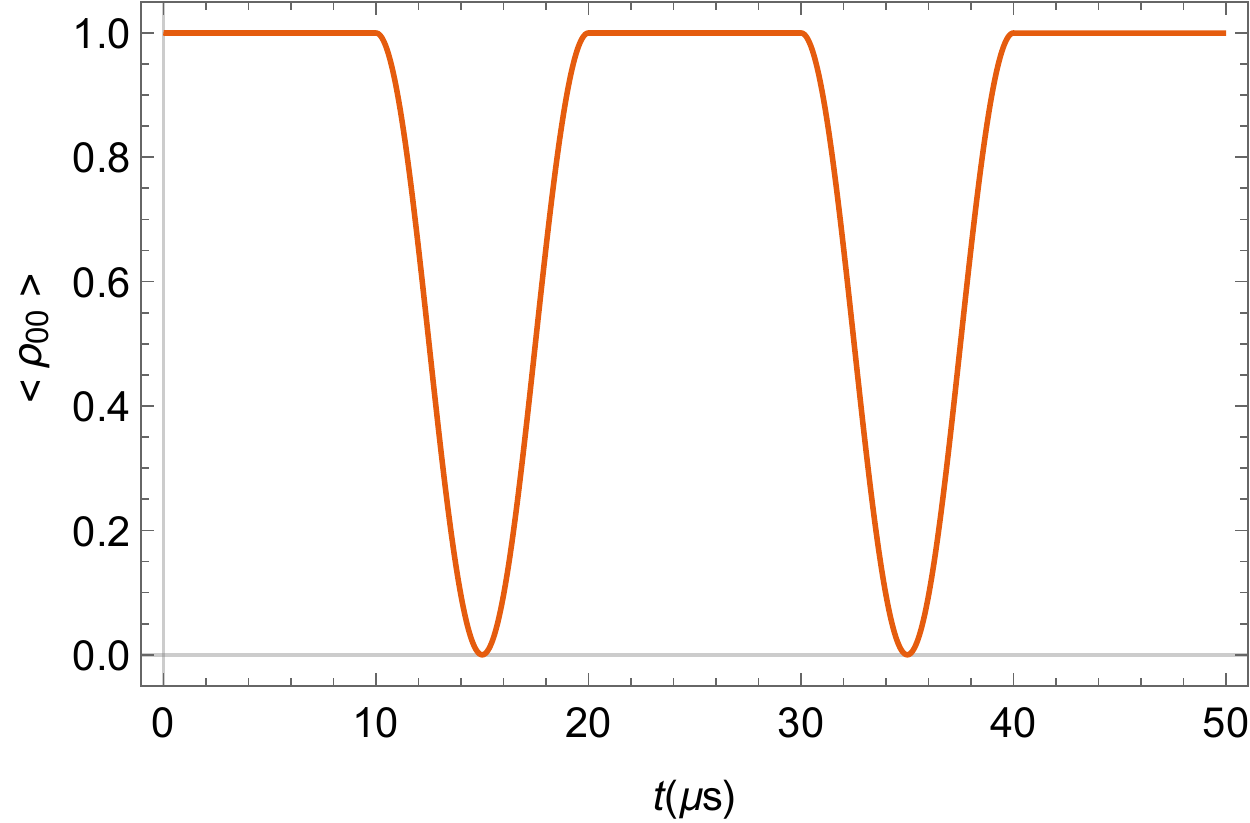}
	}
	\caption{\label{aFig1}(Color online) the dynamical evolution of $\rho_{00}(t)$. There is no attenuation in the system.}
\end{figure}

For different forms of Floquet $\mathcal{PT}$-symmetric systems, the dynamic evolution of the system are quite different. For the Hamiltonian $H_{eff}= -i \gamma(t) |1\rangle\langle 1| + \frac{\Omega(t)}{2} ( |0\rangle\langle 1| + |1\rangle\langle 0|)$, we can choose different $\gamma(t)$ and $\mathrm{\Omega}(t)$. For example, $\gamma\left(t\right)$ is a square wave oscillating between 0 and $\gamma$ with oscillation frequency $\mathrm{\omega}$, and $\Omega\left(t\right)$ is a square wave oscillating between 0 and $\Omega$ with the oscillation frequency $\mathrm{-\omega}$. When $\mathrm{\Omega}=0.1*2\pi(Mhz), \frac{\gamma}{\mathrm{\Omega}}=2$, $\frac{2\pi\omega}{\mathrm{\Omega}}=0.5$, and initial state is $|0\rangle$, we can obtain the dynamical evolution of $\rho_{00}(t)$, as shown in FIG.~\ref{aFig1}. Evidently, this is completely different from the evolution of the measured system in the text. Therefore, by controlling the system parameters, a Floquet $\mathcal{PT}$-symmetric system can be constructed to meet the research requirements.

\subsection{A3: Experimental Setup}
In our experimental scheme, the continuous measurement model is constructed by the internal state of $^{40}\mathrm{Ca}^{+}$, and the energy level structure is shown in FIG. 1(a). The Zeeman sublevels $^{2}\mathrm{S}_{1/2}(m_{J}=-1/2)$ and $^{2}\mathrm{D}_{5/2}(m_{J}=+1/2)$ of the $^{40}\mathrm{Ca}^{+}$ in a 5.2 G-magnetic field are chosen as the quantum states $|0\rangle$ and $|1\rangle$. The lifetime of the excited state $^{2}\mathrm{D}_{5/2}(m_{J}=+1/2)$ is $1.168\pm0.007 s$. $^{2}\mathrm{D}_{5/2}(m_{J}=+1/2)$ is excited to $^{2}\mathrm{P}_{3/2}(m_{J}=+3/2)$ by the circularly polarized 854\,nm laser beam. Experimentally, because the polarization of the 854\,nm laser is not perfectly circularly polarized, electrons may be excited from $^{2}\mathrm{D}_{5/2}(m_{J}=+1/2)$ to other Zeeman sublevels of $^{2}\mathrm{P}_{3/2}$. Since the lifetime of $^{2}\mathrm{P}_{3/2}(m_{J}=+3/2)$ is $6.924\pm0.019 ns$, the electron cannot exist stably and will emit spontaneously. According to the selection rule and branching ratio of the spontaneous emission, the electron has about 94\% chance to transition to $^{2}\mathrm{S}_{1/2}(m_{J}=+1/2)$ and 6\% chance to transition to $^{2}\mathrm{D}_{5/2}(m_{J}=+1/2)$. The spontaneous emission of $^{2}\mathrm{P}_{3/2}(m_{J}=+3/2)$ ensures that most of the population flows into $^{2}\mathrm{S}_{1/2}(m_{J}=+1/2)$ as the environment and not back into $^{2}\mathrm{S}_{1/2}(m_{J}=-1/2)$ as the system. After considering the polarization of 854\,nm laser and branch ratio of the spontaneous emission, the probability of this emission from $^{2}\mathrm{P}_{3/2}(m_{J}=+3/2)$ state to $^{2}\mathrm{S}_{1/2}(m_{J}=+1/2)$ is about 90\%. These experimental setup and optimizations ensure the rationality of the theoretical approximation in this paper.

$^{2}\mathrm{S}_{1/2}(m_{J}=-1/2)$ and $^{2}\mathrm{D}_{5/2}(m_{J}=+1/2)$ are coupled by the 729\,nm laser, and $^{2}\mathrm{D}_{5/2}(m_{J}=+1/2)$ and $^{2}\mathrm{P}_{3/2}(m_{J}=+3/2)$ are coupled by the 854\,nm laser. The 729\,nm and 854\,nm lasers resonate with the corresponding transition energy level, respectively. Thus, in the interaction picture, the Hamiltonian of the system can be written as:
\begin{eqnarray}
	H_{eff} = -i \Gamma |P\rangle\langle P| &+& \frac{\Omega}{2} ( |0\rangle\langle 1| + |1\rangle\langle 0|)  \nonumber\\
	&+&\frac{\Omega'}{2} ( |P\rangle\langle 1| + |1\rangle\langle P|).  \label{aeq16}
\end{eqnarray}
where $\Omega$ is the coupling strength between $|0\rangle$ and $|1\rangle$, and $\Omega'$ is the coupling strength between $|1\rangle$ and $|P\rangle$. The coupling strength can be controlled by adjusting the laser intensity. There is an electric quadrupole transition between $|0\rangle$ and $|1\rangle$, and an electric dipole transition between $|1\rangle$ and $|P\rangle$; thus, in general $\Omega'>>\Omega$ (In the experiments, $\Omega\simeq 0.1(MHz)$, and $\Omega'$ is in the order of $MHz$). $\Gamma\simeq 22 (MHz)$ is the natural linewidth of $^{2}\mathrm{P}_{3/2}(m_{J}=+3/2)$. According to the method given in literature \cite{Facchi2009,2006Continuous}, when $\Omega',\Gamma>>\Omega $, Eq.~(\ref{aeq16}) can be written as:
\begin{eqnarray}
	H_{eff} = -i \gamma |1\rangle\langle 1| + \frac{\Omega}{2} ( |0\rangle\langle 1| + |1\rangle\langle 0|).  \label{aeq17}
\end{eqnarray}
Where $\gamma=\Omega'^{2}/(2\Gamma)$. So we have the desired Hamiltonian for the $^{40}\mathrm{Ca}^{+}$ system.

In the experimental scheme, the initial state of the measured system is prepared to $^{2}\mathrm{S}_{1/2}(m_{J}=-1/2)$, i.e., state $|0\rangle$. The motion state of the $^{40}\mathrm{Ca}^{+}$ is cooled to the ground state by a series of laser cooling techniques, including the Doppler cooling, EIT cooling, and sideband cooling. The average phonon number of an ion is about 0.02. After the cooling completion, the electrons are manipulated by a laser into an $^{2}\mathrm{S}_{1/2}(m_{J}=-1/2)$. In the experiment, the fidelity of the initial state is about 99.8\%.
\end{document}